\documentclass[fleqn,12pt,twoside]{article}
\usepackage[headings]{espcrc1}

\readRCS
$Id: espcrc1.tex,v 1.2 2004/02/24 11:22:11 spepping Exp $
\usepackage{graphicx,wrapfig}
\usepackage[figuresright]{rotating}

\newcommand{\AmS}{{\protect\the\textfont2
  A\kern-.1667em\lower.5ex\hbox{M}\kern-.125emS}}

\title{3-D hydro + cascade model at RHIC 
       }

\author{Chiho Nonaka\address[DUKE]{Department of Physics,  
        Duke University, Durham, NC 27708}
\thanks{Present address: School of Physics and Astronomy, 
University of Minnesota, Minneapolis, MN 55455. 
       } and 
        Steffen A. Bass\addressmark[DUKE]
	\thanks{This work was supported by DOE grants DE-FG02-05ER41367 
	and DE-FG02-03ER41239. } 
      } 
\runtitle{3-D hydro + cascade model at RHIC}
\runauthor{C. Nonaka and S. A. Bass}

\begin{document}

% typeset front matter
\maketitle

\begin{abstract}
We present a 3-D hydro + cascade model, in which viscosity  
 and a realistic freezeout process for the hadronic phase are taken into account.  
We compare our results to experimental data and
discuss the final state interaction effects on physical 
observables. 

\end{abstract}
\section{FREEZEOUT PROCESS AND VISCOSITY IN HYDRODYNAMICS}
Hydrodynamic models have been very successful in describing the 
collective behavior of matter at RHIC, such 
as single particle spectra and elliptic flow. 
In particular the strong elliptic flow which, for the first time,  
reaches the hydrodynamic limit at   
RHIC, provides us with a new understanding of the nature of the quark-gluon plasma (QGP)
created at RHIC as strongly interacting or correlated QGP \cite{GyMc}.  
However there exist a number of experimental observations that contradict    
perfect hydrodynamic models:  transverse momentum spectra above 
2 GeV, elliptic flow at large pseudo-rapdities $\eta$ and Hanbury Brown - Twiss (HBT)
 interferometry. 
These observations suggest that there exist limitations to the application of a simple 
perfect hydrodynamic model to RHIC physics and that an improvement on a perfect   
hydrodynamic model is needed in order to obtain a
 comprehensive and unified description of the data     
from the point of view of hydrodynamics. 

In general,  hydrodynamic models require initial conditions, 
an equations of state (EoS) and freezeout conditions as parameters 
besides the relativistic hydrodynamic equation.  
One of the main advantages of the hydrodynamic model lies in its ability to 
investigate the relation between the EoS and 
physical observables via comparison to experimental data. 
However, it has been pointed out repeatedly, e.g. by Hirano \cite{Hirano},     
that details of the treatment of the freeze-out process 
can have large effects on physical observables.      
For example, the data on $P_T$ spectra and elliptic flow show 
that the assumption of chemical equilibrium \cite{Ho} or
partial chemical equilibrium \cite{PCE} in the freezeout process   
is not realistic \cite{Hirano}. 
In addition, studies of collective flow at AGS, SPS and RHIC 
based on both of hydrodynamic models and cascade models suggest that 
the effects of viscosity are not negligible \cite{St}.  
Therefore we construct a hybrid 3-D hydro + 
cascade model to include a realistic treatment of the freezeout process 
and viscosity in the hadronic phase. Such hybrid models have been implemented
in the past, however with reduced dimensionality in the hydrodynamic component
and thus with restrictions to observables at mid-rapidity \cite{hybrid}.          
As a cascade model we use UrQMD in which   
final state interactions are included correctly \cite{UrQMD_resonance}. 

%%%%%%%%%%%%%%%%%%%%%%%%%%%%%%%%%%%%%%%%%%%%%%%%%%%%%%%%%%%%%%%%%%%%%%%
\section{3-D HYDRO + CASCADE MODEL}
\subsection{3-D hydrodynamic model}
We solve the relativistic hydrodynamic equation,  
\begin{equation}
\partial_\mu T^{\mu \nu} = 0, 
\end{equation}
where $T^{\mu \nu}$ is energy momentum tensor, including  
baryon number conservation, \\   
$\displaystyle 
\partial_\mu(n_B(T,\mu)u^\mu)=0$. 
We modify our original code for the hydrodynamic expansion in 
Cartesian coordinates \cite{NoHoMu} 
to that in light-cone coordinates $(\tau,x,y,\eta)$ ($\tau=\sqrt{t^2-z^2}$), 
 in order to optimize our hydrodynamic model for ultra-relativistic high energy heavy 
collisions \cite{NoBa}.  
We employ Lagrangian hydrodynamics which   
has the following advantages over Eulerian 
hydrodynamics \cite{NoHoMu}: 
1) We can trace the adiabatic path of each volume element of fluid 
on phase diagram. 
2) We can easily investigate effects of phase transition on 
physical observables. 

We assume that hydrodynamic expansion starts at 
$\tau_0=0.6$ fm. Initial energy density and 
baryon number density are parameterized by      
\begin{eqnarray}
\epsilon(x,y,\eta)& =& \epsilon_{\rm max}W(x,y;b)H(\eta), 
\nonumber \\
n_B(x,y,\eta)& = & n_{B{\rm max}}W(x,y;b)H(\eta), 
\end{eqnarray}
where $b$ and  $\epsilon_{\rm max}$ ($n_{B{\rm max}}$) are  
the impact parameter and the maximum value of energy density 
(baryon number density), respectively.  
$W(x,y;b)$ is given by a combination of wounded nuclear model and 
binary collision model \cite{KoHeHuEsTu} and  $H(\eta)$ is given  
by $\displaystyle 
H(\eta)=\exp \left [ - (|\eta|-\eta_0)^2/2 \sigma_\eta^2 \cdot  
\theta ( |\eta| - \eta_0 ) \right ]$. 
Parameters $\epsilon_{\rm max}$, $n_{B {\rm max}}$, $\eta_0$ and  
$\sigma_\eta$ are listed in Table \ref{table:initial}. 
We set initial flow in the longitudinal direction 
to $v_L=\eta$ (Bjorken's solution) 
and $v_T=0$ in the transverse plane. 
We use an equation of state with a 1st order phase transition, namely a 
Bag model EoS with and excluded volume correction \cite{RiGo}. 
The thermal freezeout temperature is 110 MeV. 

\subsection{3-D hydro + cascade model}
The key improvements of the hybrid model approach compared to the 3-D hydrodynamic
model are the inclusion of 
viscosity in the hadronic phase as well a realistic treatment of
final state interactions  
and a species dependent freezeout process.    
For the transition from hydrodynamic model to UrQMD  
we introduce the switching temperature $T_{\rm SW}$ which  
should be chosen below but near the critical temperature ($T_{\rm c}=160$ MeV 
\cite{hybrid}).  
We set the switching temperature to 150 MeV. 
The calculated procedure is as follows: 
first hadron distributions from the hydrodynamic model are 
calculated at the switching temperature via the Cooper-Frye formula \cite{CoFr}. 
Second, initial conditions for UrQMD on an event by event basis are  
generated from the hadron distributions via the hydrodynamic expansion through a 
Monte Carlo implementation. 
%of the Cooper-Frye formula. 
Finally the UrQMD calculation starts with these 
initial conditions.    
In this scheme we neglect the reverse contribution from UrQMD to the hydrodynamic model.  
For an initial condition for 3-D hydro + UrQMD we assume the 
same parameterization as that of pure hydrodynamic model except the 
value of maximum energy density, which we have adjusted to achieve a better
agreement to experimental data. The parameters are summarized in Table 1.  
%%%%%%%%%%%%%%%%%%%%%%%%%%%%%%%%%%%%%%%%%%%%%%%%%%%%%%%%%%%%%%%%%%%%%%
\begin{table}[htb]
\vspace*{-0.6cm}
\caption{Parameters for initial conditions of pure hydro and hybrid model.}
\label{table:initial}
\begin{tabular}{|c|c|c|c|c|c|}
\hline \hline
  & $\tau_0$ (fm/$c$)  & max. of $\epsilon$ (GeV/fm$^3$) & 
  max. of $n_{B {\rm max}}$ (fm$^{-3}$) & $\eta_0$ & $\sigma_\eta$  \\ \hline
 pure hydro & 0.6    & 55   & 0.15   & 0.5   & 1.5  \\ \hline
 hydro + UrQMD & 0.6   & 43    &  0.15  & 0.5   & 1.5  \\ \hline
\end{tabular}
\vspace*{-0.8cm}
\end{table}
%%%%%%%%%%%%%%%%%%%%%%%%%%%%%%%%%%%%%%%%%%%%%%%%%%%%%%%%%%%%%%%%%%%%%%

%%%%%%%%%%%%%%%%%%%%%%%%%%%%%%%%%%%%%%%%%%%%%%%%%%%%%%%%%%%%%%%%%%%%%%
\section{RESULTS and DISCUSSIONS}
\begin{figure}[htb]
\vspace*{-1.2cm}
\begin{minipage}[t]{75mm}
\includegraphics[width=1.0\linewidth]{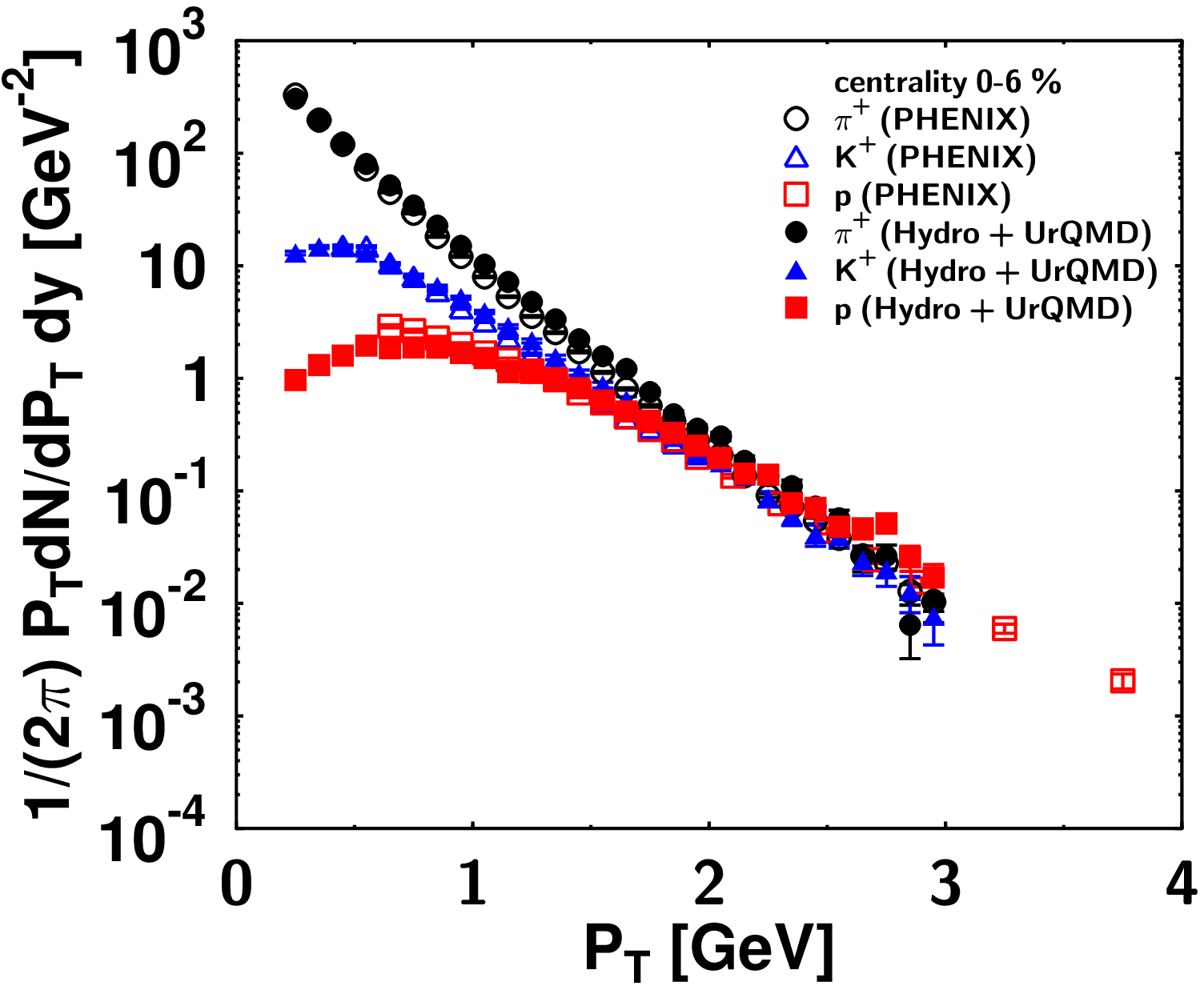}
\vspace*{-1.5cm}
\caption{$P_T$ spectra for $\pi^+$, $K^+$, $p$ at 
central collisions with PHENIX data \cite{PHENIX}.
}
\label{Fig:hydro-urqmd-pt}
\end{minipage}
\hspace{\fill}
\begin{minipage}[t]{75mm}
\includegraphics[width=1.0\linewidth]{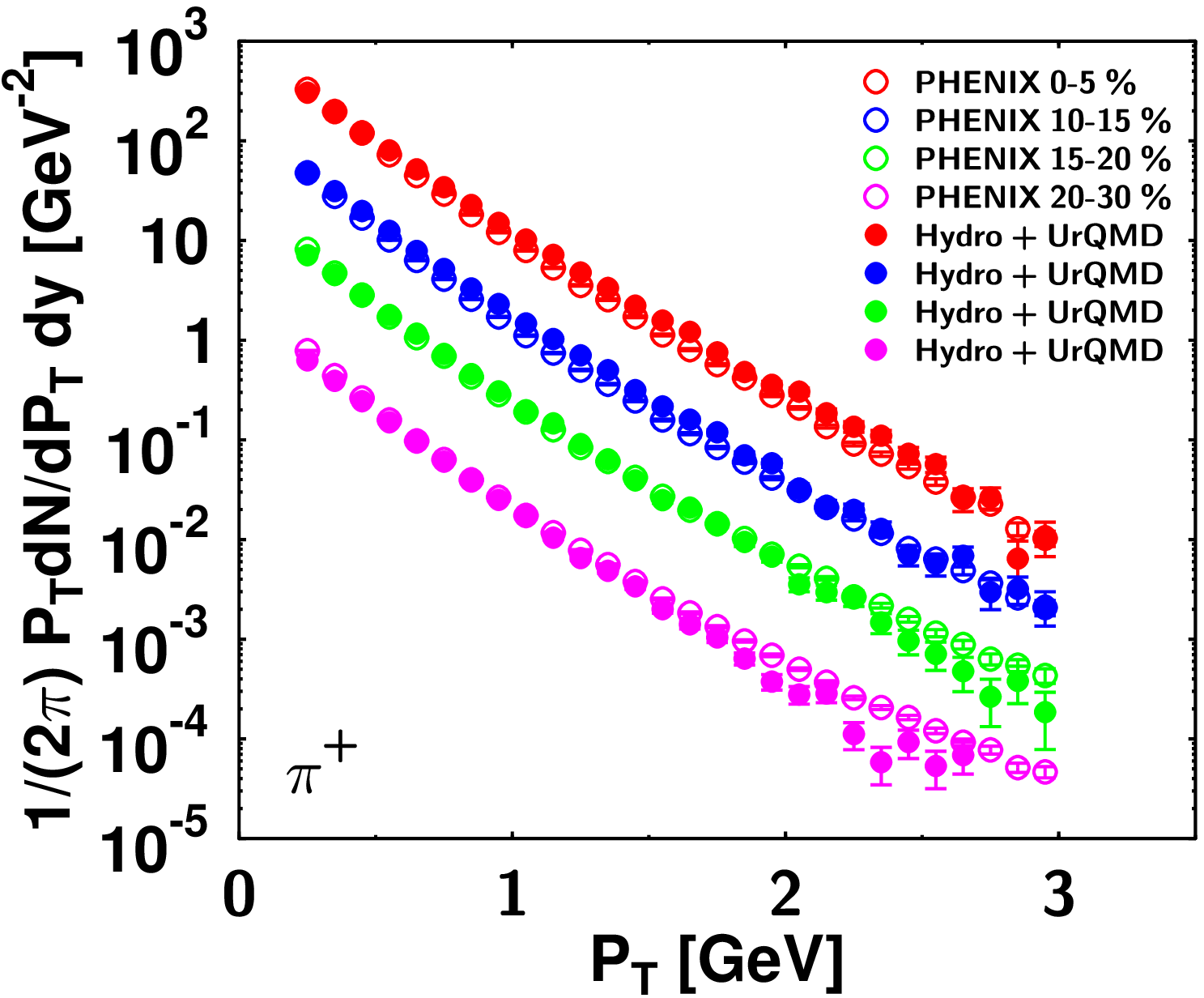}
\vspace*{-1.5cm}
\caption{Centrality dependence of $P_T$ spectra of $\pi^+$ with 
PHENIX data \cite{PHENIX}.
}
\label{Fig:hydro-urqmd-pt-cent}
\end{minipage}
\vspace*{-0.8cm}
\end{figure}

Figure \ref{Fig:hydro-urqmd-pt} shows the $P_T$ spectra of $\pi$, $K$, and 
$p$ at Au + Au $\sqrt{s}=200$ GeV central collisions 
based on 3-D hydro + UrQMD. 
Our calculated results are very close to PHENIX data \cite{PHENIX} 
up to $P_T \sim 2$ GeV.   
In 3-D hydro + UrQMD model the hadron ratio $\pi/p$ and  $\pi/K$ are 
obtained correctly. In the case of the pure 3-D hydrodynamic model 
with low thermal freezeout temperature under the assumption of 
chemical equilibrium, we fail to obtain the correct hadron 
ratios \cite{NoBa}. 
%In other words, for $K$ and $p$ spectra renormalization to 
%the ratio at chemical freezeout is needed \cite{NoBa}. 
This observation emphasizes the importance of the realistic treatment for 
the freezeout process, but further studies are necessary 
to investigate the effects of cross-over transition as well. 

Figure \ref{Fig:hydro-urqmd-pt-cent} shows the centrality 
dependence of $P_T$ spectra of $\pi^+$. 
We can see that in peripheral collisions the difference 
between experimental results and calculated results 
appears at lower $P_T$ compared to central collisions.  
These deviations are indicative of the diminished importance
of the soft, collective, physics described by hydrodynamics compared to
the contribution of jet-physics in these peripheral events.

%%%%%%%%%%%%%%%%%%%%%%%%%%%%%%%%%%%%%%%%%%%%%%%%%%%%%%%%%%%%%%%%%%%%
%%%%%%%%%%%%%%%%%%%%%%%%%%%%%%%%%%%%%%%%%%%%%%%%%%%%%%%%%%%%%%%%%%%%

%%%%%%%%%%%%%%%%%%%%%%%%%%%%%%%%%%%%%%%%%%%%%%%%%%%%%%%%%%%%%%%%%%%%
\begin{figure}[htb]
\vspace*{-1.2cm}
\begin{minipage}[t]{75mm}
\includegraphics[width=1.0\linewidth]{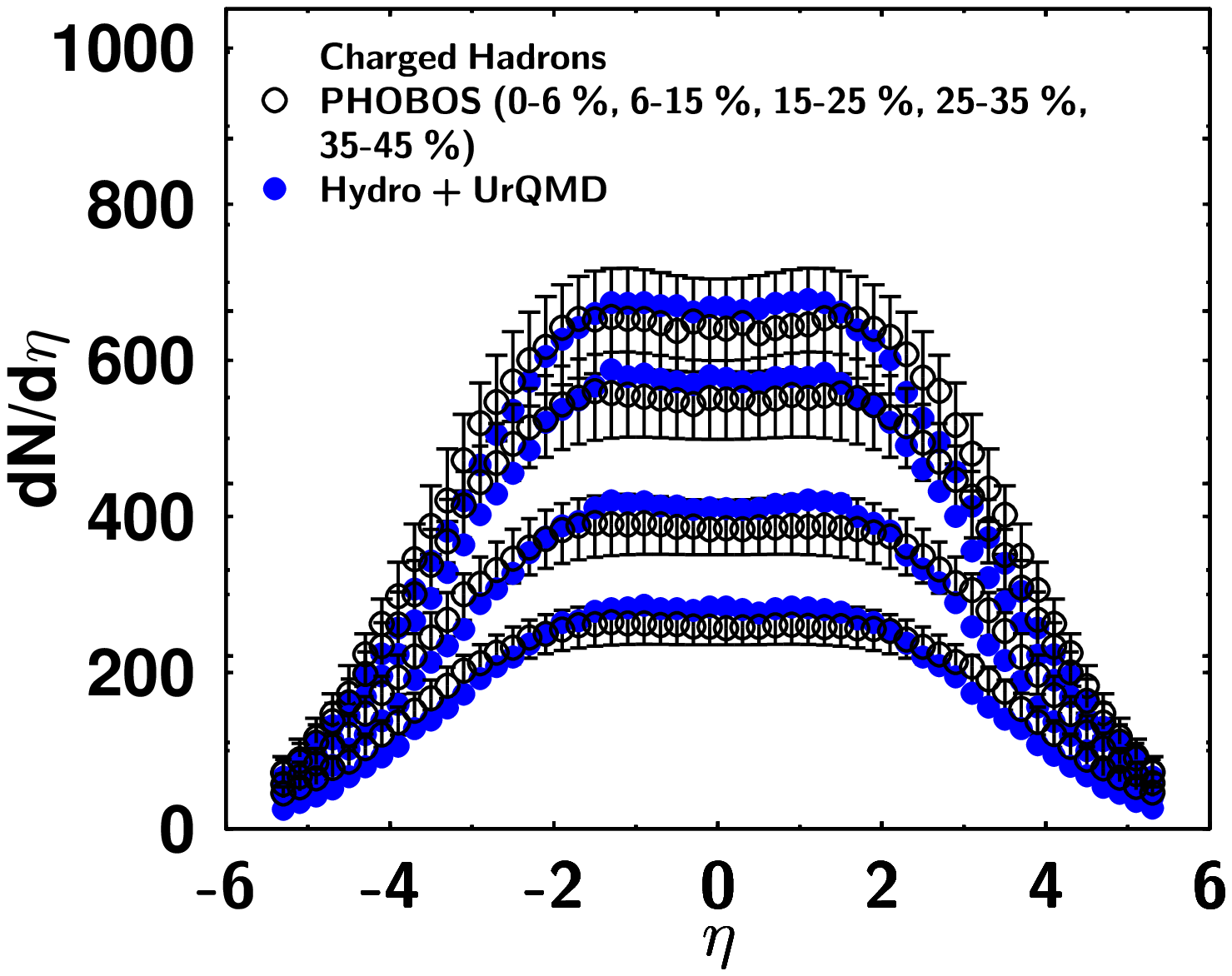}
\vspace*{-1.5cm}
\caption{Centrality dependence of rapidity distribution  
with PHOBOS data \cite{PHOBOS_rap}. 
}
\label{Fig:eta_b}
\end{minipage}
\hspace{\fill}
\begin{minipage}[t]{75mm}
\includegraphics[width=1.0\linewidth]{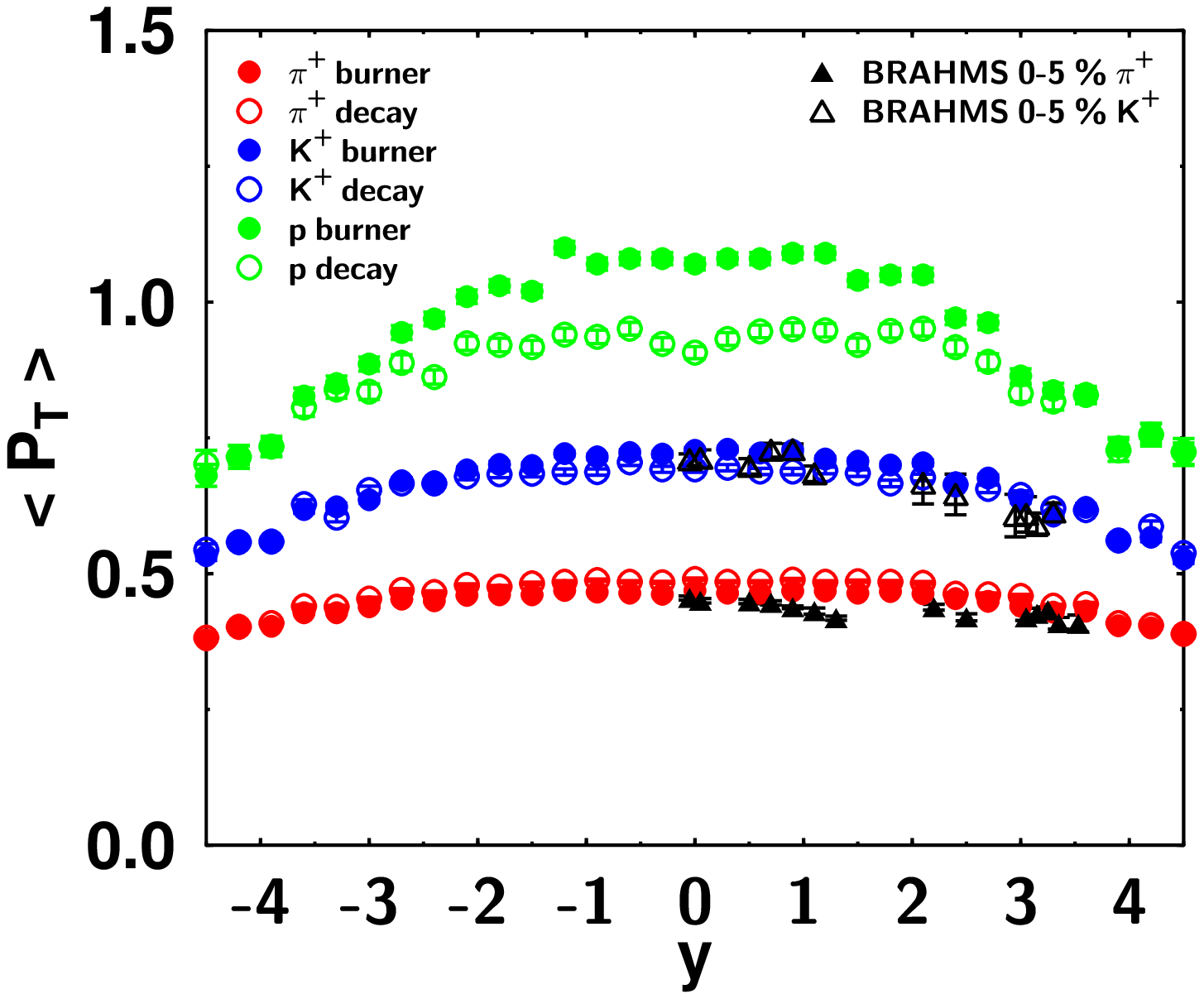}
\vspace*{-1.5cm}
\caption{Mean $P_T$ for $\pi^+$, $K^+$, $p$ as a function of $y$  
with BRAHMS data \cite{BRAHMS}. }  
%Open symbols: from decay, Solid symbols: from burner}
\label{Fig-meanpt}
\end{minipage}
\vspace*{-0.9cm}
\end{figure}
%%%%%%%%%%%%%%%%%%%%%%%%%%%%%%%%%%%%%%%%%%%%%%%%%%%%%%%%%%%%%%%%%%%%
Figure \ref{Fig:eta_b} shows the centrality dependence of the
pseudorapidity distribution of charged hadrons together with PHOBOS 
data \cite{PHOBOS_rap}. The impact parameters are set to $b=2.4, 
4.5, 6.3$ and $7.9$ fm for 0-6 \%, 6-15 \%, 15-25 \% and 25-35 \%, 
respectively.      
Our results agree with the experimental data not only at mid rapidity  
but also at large rapidity.   

Figure \ref{Fig-meanpt} shows the mean $P_T$ as a function of  
rapidity for $\pi$, $K$ and $p$. The difference between 
open symbols (hadronic decays without rescattering) and solid symbols (full hadronic
rescattering)  
determines the reaction dynamics of the final state interactions. 
We find that the average transverse momentum $\langle  P_T \rangle$ for $\pi$ 
decreases whereas the proton $P_T$ increases. Since the number of 
$\pi$ is larger compared to that of $p$, the effect on the protons per particle 
is larger than for the pions. This transfer of transverse momentum is commonly
referred to as the "pion-wind".  

In summary, we have presented a fully 3-D implementation of a hybrid
hydro+micro transport model. The model includes a realistic treatment
of hadronic freeze-out as well as an implementation of viscous effects
in the hadronic phase. We have compared single-particle spectra and
pseudo-rapidity distributions to data and have studied the reaction
dynamics of protons, pions and kaons in the hadronic phase of heavy-ion
collisions at RHIC.

%Furthermore the success of this hybrid model of 3-D hydro + UrQMD 
%model to explanation of various experimental data at RHIC suggests that 
%necessity of applying a suitable and realistic model to each process  
%in heavy ion collisions, which helps us to understand  
%RHIC physics comprehensively. 

%%%%%%%%%%%%%%%%%%%%%%%%%%%%%%%%%%%%%%%%%%%%%%%%%%%%%%%%%%%%%%%%%%%%%
%\begin{wrapfigure}{l}{75mm}
%\begin{flushright}
%\vspace*{-0.9cm}
%\includegraphics*[width=1.\linewidth]{peta_b.eps}
%\caption{
%}
%\vspace*{-1.1cm}
%\label{Fig}
%\end{flushright}
%\end{wrapfigure}
%%%%%%%%%%%%%%%%%%%%%%%%%%%%%%%%%%%%%%%%%%%%%%%%%%%%%%%%%%%%%%%%%%%%

\end{document}